\documentclass[12pt]{article}
\usepackage{graphicx, epstopdf, amsmath, amssymb, amsfonts, comment}

\newtheorem{theorem}{Theorem}%[section]
\newtheorem{lemma}{Lemma}%[section]
\newtheorem{remark}{Remark}%[section]

\def\R{\mathbb{R}}

\begin{document}

\title{Pricing Perpetual Put Options by the Black--Scholes Equation with a Nonlinear Volatility Function
}

\author{Maria do Ros\'ario Grossinho, Yaser Kord Faghan\thanks{Universidade de Lisboa, ISEG, Department of Mathematics, and CEMAPRE, Rua do Quelhas, 6, 1200-781 Lisboa, Portugal. {\tt mrg@iseg.ulisboa.pt}}, 
Daniel \v Sev\v{c}ovi\v{c}
\thanks{Department of Applied Mathematics and Statistics, Comenius University, 842 48 Bratislava, Slovak Republic. {\tt sevcovic@fmph.uniba.sk}}
}

\maketitle

\begin{abstract}
We investigate qualitative and quantitative behavior of a solution of the mathematical model for pricing American style of perpetual put options. We assume the option price is a solution to the stationary generalized Black-Scholes equation in which the volatility function may depend on the second derivative of the option price itself. We prove existence and uniqueness of a solution to the free boundary problem. We derive a single implicit equation for the free boundary position and the closed form formula for the option price. It is a generalization of the well-known explicit closed form solution derived by Merton for the case of a constant volatility. We also present results of numerical computations of the free boundary position, option price and their dependence on model parameters. 
\end{abstract}

\vskip8pt\noindent
{\bf Key words.} Option pricing, nonlinear Black-Scholes equation, perpetual American put option,  early exercise boundary

\vskip8pt\noindent
{\bf 2000 Mathematical Subject Classifications.} 35R35 91B28  62P05

\section{Introduction}

In a stylized financial market, the price of a European option can be computed from a solution to the well-known Black--Scholes linear parabolic equation derived by Black and Scholes in \cite{BS}, and, independently by Merton in \cite{Merton1971} (c.f. Kwok \cite{Kw}, Dewynne {\em et al.} \cite{DH}, Hull \cite{H}). A European call (put) option is the right but not obligation to purchase (sell) an underlying  asset at the expiration price $E$ at the expiration time $T$. 

In contrast to European options, American style options can be exercised anytime in the temporal interval $[0,T]$ with the specified time  of obligatory expiration at $t=T$. A mathematical model for pricing American put options leads to a free boundary problem. It consists in construction of a function  $V=V(S,t)$ together with the early exercise boundary profile $S_{f}:[0, T]\to \R$ satisfying the following conditions:

\begin{enumerate}
\item $V$ is a solution to the Black--Scholes partial differential equation:
\begin{equation}
\partial_t V + \frac{1}{2} \sigma^2 S^2 \partial^2_S V + r S \partial_S V -r V =0 
\label{nonlinear-BS}
\end{equation} 
defined on the time dependent domain  $S > S_{f}(t)$ where $0<t<T$. Here $\sigma$ is the volatility of the underlying asset price process, $r>0$ is the interest rate of a zero-coupon bond. A solution $V=V(S,t)$ represents the price of an American style put option for the underlying asset price $S>0$ at the time $t\in[0,T]$;

\item $V$ satisfies the terminal pay-off condition:
\begin{equation}
\label{put-terminal}
V(S,T)=\max(E-S,\, 0);
\end{equation}

\item
and boundary conditions for the American put option:
\begin{equation}
\label{put-b-conditions}
V(S_{f}(t), t) = E-S_{f}(t), \quad 
\partial_S V(S_f(t),t) = -1,\quad 
V(+\infty, t) = 0, 
\end{equation}
for $S=S_f(t)$  and $S=\infty$. 
\end{enumerate}

Since the seminal paper by Brennan and Schwartz \cite{Brennan1977} American style of a put option has been investigated by many authors (c.f. Kwok \cite{Kw} and references therein). Various accurate analytic approximations of the free boundary position have been derived by Stamicar, \v{S}ev\v{c}ovi\v{c} and Chadam \cite{SSC}, Evans, Kuske and Keller \cite{KK}, and by S.~P. Zhu and Lauko and \v{S}ev\v{c}ovi\v{c} in recent papers \cite{SPZHU} and \cite{LS} dealing with analytic approximations on the whole time interval. 

If the volatility $\sigma>0$  in (\ref{nonlinear-BS}) is constant then (\ref{nonlinear-BS}) is a classical linear Black--Scholes parabolic equation derived by Black and Scholes in \cite{BS}. If we assume the volatility $\sigma>0$ is a function  of the solution $V$ then  equation (\ref{nonlinear-BS}) with such a diffusion coefficient represents a nonlinear generalization of the Black--Scholes equation. 
In this paper we  focus our attention to the case when the diffusion coefficient
$\sigma^2$ may depend on the asset price $S$ and the second derivative $\partial^2_S V$ of the option price. More precisely, we will assume that 
\begin{equation}
\sigma = \sigma(S \partial^2_S V)\,,
\label{doplnky-c-sigma}
\end{equation}
i.e. $\sigma$ depends on the product $S\partial^2_S V$ of the asset price $S$ and the second derivative (Gamma) of the option price $V$. Recall that the nonlinear Black--Scholes equation (\ref{nonlinear-BS}) with a nonlinear  volatility $\sigma$ having the form of (\ref{doplnky-c-sigma}) arises from option pricing  models taking into account nontrivial transaction costs, market feedbacks and/or risk from a volatile  (unprotected) portfolio. The linear Black--Scholes equation with a constant volatility $\sigma$ has  been derived under several restrictive assumptions like e.g., frictionless, liquid and complete markets, etc. Such assumptions have been relaxed in order to model the  presence of transaction costs (see e.g., Leland \cite{Le}, Hoggard {\em et al.} \cite{HWW}, Avellaneda and  Paras
\cite{AP}), feedback and illiquid market effects due to large traders choosing given stock-trading  strategies (Frey \cite{F}, Frey and Patie \cite{FP}, Frey and Stremme \cite{FS},  Sch\"onbucher and  Wilmott \cite{SW}), imperfect replication and investor's preferences (Barles and Soner  \cite{BaSo}), risk from unprotected portfolio (Kratka \cite{Kr}, Janda\v{c}ka and \v{S}ev\v{c}ovi\v{c} \cite{JS} or \cite{Se2}). 

In the Leland model (generalized for more complex  option strategies by Hoggard \emph{et al.} \cite{HWW}) the  volatility is given by
$\sigma^2  = \sigma_0^2 ( 1 + \hbox{Le}\, \hbox{sgn}(\partial^2_S V))$ where  $\sigma_0>0$ is the constant historical volatility of the underlying asset price process and $\hbox{Le}>0$ is the so-called Leland number. Another nonlinear Black--Scholes model has been derived by Frey in \cite{F}. In this model the asset dynamics takes into account the presence of feedback effects due to a large trader choosing his/her stock-trading strategy (see also \cite{SW}). The diffusion coefficient $\sigma^2$ is again non-constant:
\begin{equation}
\sigma(S \partial^2_S V)^2 = \sigma_0^2 \left(1-\mu S\partial^2_S V\right)^{-2},
\label{doplnky-c-frey}
\end{equation}
where $\sigma_0^2, \mu>0$ are constants. 

Next example of the Black--Scholes equation with a non-constant volatility is the so-called Risk Adjusted Pricing Methodology model proposed by Kratka in \cite{Kr} and revisited by Janda\v{c}ka and \v{S}ev\v{c}ovi\v{c} in \cite{JS}. In the Risk adjusted pricing methodology model (RAPM) the purpose is to  optimize  the time-lag between consecutive portfolio adjustments in such way that the sum of the rate of transaction costs and the rate of a risk from unprotected portfolio is minimal. In this model, the volatility is again non-constant and has the form:
\begin{equation}
\sigma(S\partial^2_S V)^2 = \sigma_0^2 \left(1 + \mu (S\partial^2_S V)^{\frac{1}{3}} \right)\,,
\label{doplnky-c-RAPM}
\end{equation}
where $\sigma_0>0$ is a constant historical volatility of the asset price returns and $\mu=3(C^2 R/2\pi)^{\frac{1}{3}},$ where $C, R\ge 0$ are  nonnegative constants representing the transaction cost measure and the risk premium measure, respectively (see \cite{JS} for details). Recently, explicit solutions to the Black--Scholes equation with varying volatility of the form (\ref{doplnky-c-frey}) and (\ref{doplnky-c-RAPM}) have been derived by Bordag and Chmakova \cite{BC} and Bordag \cite{Bordag2011,Bordag2015}. 

Another important contribution in this direction has been presented by Amster, Averbuj, Mariani and Rial in \cite{AAMR}, where the transaction costs are assumed to be a non-increasing linear function of the form $C(\xi)=C_0-\kappa\xi$, ($C_0, \,\kappa >0$), depending on the volume of traded stocks $\xi\geq0$ that is needed to hedge the replicating portfolio. In the model studied by Amster \emph{et al.} \cite{AAMR}  the volatility function has the following form:

\begin{equation}\label{sigma:Amster}
\sigma(S\partial^2_S V)^2=\sigma_0^2 \left( 1 -  \mathrm{Le}\, \mathrm{sgn} \left(S\partial_S^2 V \right) + \kappa  S \partial_S^2 V \right).
\end{equation}
A disadvantage of such a transaction costs function is the fact that it may attain negative values when the amount of transactions exceeds the critical value $\xi = C_0/\kappa$. The model (\ref{sigma:Amster}) has been generalized to a class of nonnegative variable transaction cost function by \v{S}ev\v{c}ovi\v{c} and \v{Z}it\v{n}ansk\'a in  \cite{SZ2016}. 

In \cite{BaHowison} Bakstein and Howison investigated a parametrized model for liquidity effects arising from the asset trading. In their model the volatility function is a quadratic function of the term $S\partial_S^2 V$:
\begin{align}
\sigma(S\partial^2_S V)^2= & \sigma_0^2 \Biggl(  1+{\gamma}^2(1-\alpha)^2 + 2\lambda  S \partial^2_S V +  \lambda^2(1-\alpha)^2\left(S\partial^2_S V\right)^2  
\nonumber \\ 
&+ 2\sqrt{\frac{2}{\pi}} {\gamma} \, \mathrm{sgn}\left(S\partial^2_S V\right) + 2\sqrt{\frac{2}{\pi}} \lambda (1-\alpha)^2 {\gamma}  \left|S \partial^2_S V \right| \Biggr).
\end{align}
The parameter $\lambda$ corresponds to a market depth measure, i.e. it scales the slope of the average transaction price. The parameter ${\gamma} $ models the relative bid--ask spreads and it is related to the Leland number through relation $2{\gamma}\sqrt{2/\pi}=\mathrm{Le}$. Finally, $\alpha$ transforms the average transaction price into the next quoted price, $0\leq \alpha \leq 1 $. An interesting generalization of the linear Black-Scholes equation with the volatility function polynomially dependent on $S\partial^2_S V$ has been proposed by Cetin, Jarrow and Protter \cite{Cetin2004}.

Note that if additional model parameters (e.g., Le, $\mu, \kappa, \gamma, \lambda$) are vanishing then all the aforementioned nonlinear models are consistent with the original Black--Scholes equation, i.e. $\sigma=\sigma_0$. Furthermore, for call or put options, the function $V$ is convex in the $S$ variable. 

The main purpose of this paper is to investigate qualitative and quantitative behavior of a solution to the problem of pricing American style of perpetual put options. We assume the option price is a solution to a stationary generalized Black-Scholes equation with a nonlinear volatility function. We prove existence and uniqueness of a solution to the free boundary problem. We derive a single implicit equation for the free boundary position and the closed form formula for the option price. It is a generalization of the well-known explicit closed form solution derived by Merton for the case of a constant volatility. We also present results of numerical computations of the free boundary position, option price and their dependence on model parameters. In the recent paper \cite{Grossinho2017} we investigated the case when the volatility function may depend on $S$ and $\partial^2_S V$ including other models proposed by Barles and Soner \cite{BaSo}, Frey and Patie \cite{FP}, Frey and Stremme \cite{FS}. However, for these models there is no single implicit equation for the free boundary position and numerical methods have to be adopted.

The paper is organized as follows. In the next section we recall mathematical formulation of the perpetual American put option pricing model. We furthermore present the explicit solutions for the case of the constant volatility derived by Merton. In Section 3 we prove the existence and uniqueness of a solution to the free boundary problem. We derive a single implicit equation for the free boundary position $\varrho$ and the closed form formula for the option price. The first order expansion of the free boundary position with respect to the model parameter is also derived. We construct suitable sub-- and supper--solutions based on Merton's explicit solutions. In Section 4 we present results of numerical computations of the free boundary position, option price and their dependence on the model parameter. 

\section{Perpetual American put options}

In this section we analyze the problem of pricing American perpetual put options.  By definition, perpetual options are options with a very long maturity $T\to\infty$. Notice that both the option price and the early exercise boundary position depend on the remaining time $T-t$ to maturity.
Recently, stationary solutions to generalized Black--Scholes equation have been investigated by Grossinho \emph{et al.} in \cite{GM,FGS}.
Suppose that there exists a limit of the solution $V$ and early exercise boundary position $S_f$ for the maturity $T\to\infty$.

For an American style put option the limiting price $V = V(S)=\lim_{T-t\to\infty}V(S,t)$ and the limiting early exercise boundary position $\rho=\lim_{T-t\to\infty}S_f(t)$ of the  perpetual put option is a solution to the stationary nonlinear Black--Scholes partial differential equation:
\begin{equation}
\frac{1}{2} \sigma(S\partial^2_S V)^2 S^2 \partial^2_S V + r S \partial_S V -r V =0, \qquad S > \varrho,
\label{perpetual}
\end{equation}
and
\begin{equation}
 V(\varrho) = E-\varrho, \quad 
\partial_S  V (\varrho) = -1,
\quad  V(+\infty) = 0.
\label{perpetual-bc}
\end{equation}

Our purpose is to analyze the system of equations  (\ref{perpetual})--(\ref{perpetual-bc}). In what follows, we will prove the existence and uniqueness of a solution pair $(V(\cdot),\varrho)$ to (\ref{perpetual})--(\ref{perpetual-bc}).

In the rest of the paper, we will assume the volatility function 
\begin{equation}
\R_0^+ \ni H \mapsto \sigma(H)^2 \in \R_0^+ 
\label{nondecreasing}
\end{equation}
is non-decreasing, $\sigma(0)>0$ and such that the function $H \mapsto \sigma(H)^2 H$ is $C^1$ smooth for $H\ge 0$. Under these assumptions there exists an increasing inverse function  $\beta:\R_0^+\to \R_0^+$ such that 
\begin{equation}
\frac12 \sigma(H)^2 H  = u \qquad\hbox{iff}\qquad H= \beta(u).
\label{betafunction}
\end{equation}
which is an $C^1$ continuous and non-decreasing function such that $\beta(0) =0$, and $\beta(u)>0$ for $u>0$. As $u=\frac12 \sigma(\beta(u))^2 \beta(u) \ge \frac12 \sigma(0)^2 \beta(u)$ we have
\begin{equation}
\beta(u)\le M_1 u \quad\hbox{for all}\ u\ge 0,
\label{upperbound}
\end{equation}
where $M_1=2/\sigma(0)^2$. Moreover, for any $U_0>0$ there exists $M_0>0$ such that 
\begin{equation}
\beta(u)\ge M_0 u \quad\hbox{for all}\ u\in[0,U_0].
\label{lowerbound}
\end{equation}

Notice that the transformation $H=S\partial_S^2 V$ is a useful tool when analyzing nonlinear generalizations of the Black--Scholes equations. For  example, using this transformation the fully nonlinear Black--Scholes equation with a volatility function $\sigma=\sigma(S\partial_S^2 V)$ can be transformed into a quasilinear equation for the new variable $H$ (see \cite{JS} and \cite{SeA} for details).

\subsection{The Merton explicit solution for the constant volatility case}
\label{sec-merton}

\begin{figure}
\begin{center}
\includegraphics[width=0.45\textwidth]{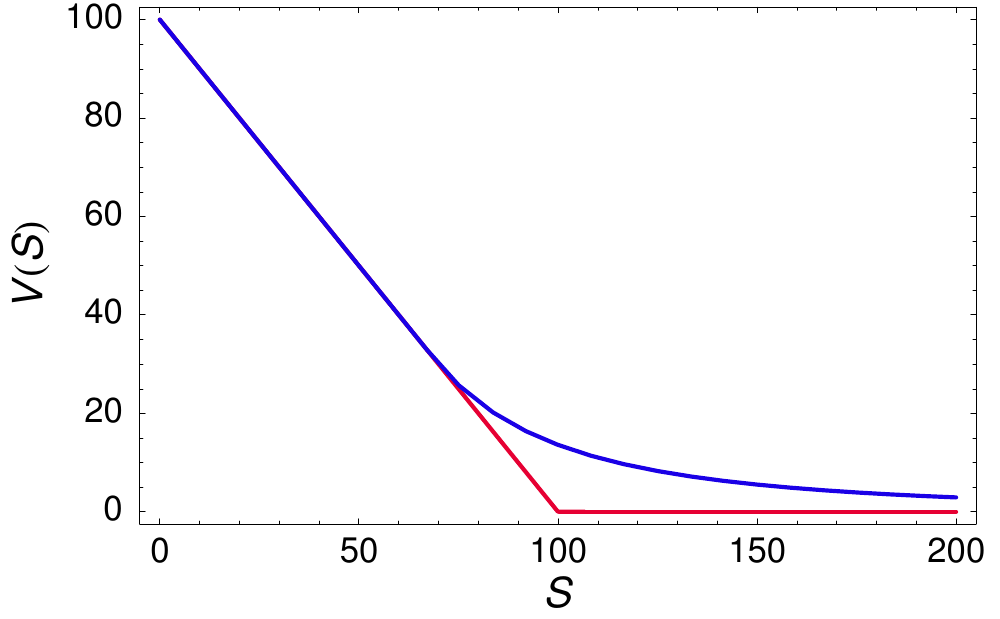}
\end{center}

\caption{A plot of the price $V(S)$ of a perpetual American put option and the pay-off diagram $\max(E-S,0)$ for the parameters: $E=100,  r=0.1$
and constant volatility $\sigma_0=0.3$ and $\gamma=2r/\sigma_0^2$.}
\label{obr-amer-perpetual}
\end{figure}

In the case of a constant volatility $\sigma\equiv \sigma_0$ the free boundary value problem  (\ref{perpetual})--(\ref{perpetual-bc}) for the function $V$ and the limiting early exercise boundary position $\varrho$ has a simple explicit solution discovered by Merton \cite{Merton1973}. The closed form solution has the following form:
\begin{equation}
V(S) = \left\{
\begin{array}{ll}
 \frac{E}{1+\gamma} \left(\frac{S}{\varrho}\right)^{-\gamma}, & S> \varrho, \\ 
 E-S,  &  0<S\le \varrho,
\end{array} 
\right.
\label{explicit-solution}
\end{equation}
where 
\begin{equation}
\varrho=E \frac{\gamma}{1+\gamma}, \qquad \gamma = \frac{2r}{\sigma_0^2}.
\label{gamma}
\end{equation}
A graph of a perpetual American put option with the constant volatility is shown in Fig.~\ref{obr-amer-perpetual}.

\section{Existence and uniqueness of solutions}

In this section we will focus our attention on existence and uniqueness of a solution to the problem (\ref{perpetual})--(\ref{perpetual-bc}). 

\subsection{Explicit formula for the perpetual American put option price}

Since $\beta$ is the inverse function to $\frac12 \sigma(H)^2 H$ the pair $(V(\cdot),\varrho)$ is a solution to (\ref{perpetual}) if and only if 
\[
 S\partial_S^2 V = \beta ( r V/S - r \partial_S V).
\]
Let us introduce the following transformation of variables:
\begin{equation}
U(x) = r \frac{V(S)}{S} - r \partial_S V(S)= - r S\partial_S\left(\frac{V(S)}{S}\right), \qquad\hbox{where} \quad x=\ln S .
\label{transformation}
\end{equation}
Since 
\[
 \partial_x U(x) = \partial_S \left( r V(S)/S - r \partial_S V \right) \frac{d S}{d x} = 
 -r S\partial_S^2 V(S) + r S \partial_S\left(\frac{V(S)}{S}\right) 
\]
the function $U(x)$ is a solution to the initial value problem 
\begin{eqnarray}
&& \partial_x U(x) = - U(x) - r \beta(U(x)), \qquad x> x_0=\ln \varrho, \label{Ueq}
\\
&& U(x_0) = \frac{r E}{\varrho}. \label{Uinit}
\end{eqnarray}
The initial condition (\ref{Uinit}) easily follows from the smooth pasting conditions $V(\varrho)=E-\varrho$ and $\partial_S V(\varrho) =-1$. Equation (\ref{Ueq}) can be easily integrated. We have the following result:

\begin{lemma}\label{lemma-1}
A solution $U=U(x)$ to the initial value problem (\ref{Ueq})-(\ref{Uinit}) is uniquely given by
\[
 U(x) = G^{-1} ( - x + x_0),\quad\hbox{for}\ \ x>x_0=\ln\varrho, 
\]
where
\begin{equation}\label{funG}
 G(U) = \int_{U(x_0)}^U \frac {1}{u + r \beta(u)} du.
\end{equation}
\end{lemma}

Taking into account the estimates (\ref{upperbound}) and (\ref{lowerbound}) we can summarize the useful properties of the function $G$:

\begin{lemma}\label{lemma-2}
The function $G:\R_0^+\to \R$ is non-decreasing and $G^{-1}(0) = r E/\varrho$. Furthermore,  $G(+\infty) = +\infty$, $G(0) = -\infty$, and, consequently, $G^{-1}(-\infty)=0$. 
\end{lemma}

Since
\[
- r S\partial_S\left(\frac{V(S)}{S}\right) = U(\ln S) = G^{-1} ( - \ln S + \ln \varrho)
\]
by taking into account the boundary condition $V(+\infty) = 0$ we conclude that the solution to equation (\ref{perpetual}) is given by
\[
V(S) = \frac{S}{r} \int_S^\infty  G^{-1} \left(- \ln\left(\frac{s}{\varrho}\right)\right) \frac{d s}{s}.
\]
Using the substitution $u=G^{-1}(-\ln(s/\varrho))$ we have 
\[
\frac{d s}{s} = - G'(u) du = - \frac{1}{u+r \beta(u)} du. 
\]
As $G^{-1}(-\infty) = 0$ the expression for $V(S)$ can be simplified as follows:
\begin{equation}
V(S) = \frac{S}{r} \int_0^{G^{-1}(-\ln(S/\varrho))} \frac{u}{u+r\beta(u)} d u.
 \label{Vexpression}
\end{equation}

\subsection{Equation for the free boundary position}

Using the expression (\ref{Vexpression}) we can derive a single implicit integral equation for the free boundary position $\varrho$. Clearly, $V(\varrho) = E-\varrho$ if and only if
\begin{equation}
E - \varrho = \frac{\varrho}{r} \int_0^{G^{-1}(0)} \frac{u}{u+r\beta(u)} d u.
\label{Veq}
\end{equation}
As $G^{-1}(0)= \frac{r E}{\varrho}$ we obtain
\begin{equation}
 \frac{r E}{\varrho} = r + \int_0^{\frac{r E}{\varrho}} \frac{u}{u+r\beta(u)} d u
 = r + \frac{r E}{\varrho} - r\int_0^{\frac{r E}{\varrho}} \frac{\beta(u)}{u+r\beta(u)} d u
\label{rhoeq}
\end{equation}
Therefore the free boundary position $\varrho$ is a solution to the following implicit equation:
\[
  \int_0^{\frac{r E}{\varrho}} \frac{\beta(u)}{u+r\beta(u)} d u =1.
\]

\subsection{Main result}
In this section we summarize the previous results and state the main result on existence and uniqueness of a solution to the perpetual American put option pricing problem (\ref{perpetual})--(\ref{perpetual-bc}). 

\begin{theorem}\label{th-1}
Suppose that the volatility function $\sigma:\R_0^+ \to \R^+$ is non-decreasing, $\sigma(0)>0$ and such that the function $H \mapsto \sigma(H)^2 H$ is $C^1$ smooth for $H\ge 0$.

Then the perpetual American put option problem (\ref{perpetual})--(\ref{perpetual-bc}) has a unique solution $(V(\cdot),\varrho)$ where the free boundary position $\varrho$ is a solution to the implicit equation
\begin{equation}
 \int_0^{\frac{r E}{\varrho}} \frac{\beta(u)}{u+r\beta(u)} d u =1,
\label{theo-rho}
\end{equation}
and the option price $V(S)$ is given by 
\begin{equation}
V(S) = \frac{S}{r} \int_0^{G^{-1}(-\ln(S/\varrho))} \frac{u}{u+r\beta(u)} d u,
\label{theo-V}
\end{equation}
where $\beta$ is the inverse function to the function $H\mapsto \frac12 \sigma(H)^2 H$.
\end{theorem}

\noindent {P r o o f.} According to results in Section 3.2 it suffices to prove that (\ref{theo-rho}) has the unique solution $\varrho$. To this end, let us introduce the auxiliary function: 
\[
 \phi(y) = \int_0^{y} \frac{\beta(u)}{u+r\beta(u)} d u
\]
we have $\phi'(y) >0, \phi(0)=0$. For a fixed $U_0>0$ we have $\beta(u)\ge \beta(U_0) >0$ for $u\ge U_0$, and 
\[
 \phi(+\infty) = \int_0^{\infty} \frac{\beta(u)}{u+r\beta(u)} d u\ge
\int_{U_0}^{\infty} \frac{\beta(u)}{u+r\beta(u)} d u
\ge \frac{\beta(U_0)}{1+ r M_1}\int_{U_0}^{\infty} \frac{1}{u} d u
=+\infty.
\]
Hence equation (\ref{theo-rho}) has the unique solution $\varrho>0$.  Clearly, $\varrho<E$ because the right hand side of (\ref{Veq}) is positive. 

Since $\varrho$ is a solution to (\ref{rhoeq}) we have $V(\varrho) = E-\varrho$. Moreover, as
\[
 U(x) = r \frac{V(S)}{S} - r \partial_S V(S),\qquad\hbox{where} \quad x=\ln S
\]
(see (\ref{transformation})) we obtain, for $x_0=\ln \varrho$, 
\[
\partial_S V(\varrho)=\frac{V(\varrho)}{\varrho} - \frac{U(x_0)}{r}
=\frac{E-\varrho}{\varrho} -  \frac{E}{\varrho} = -1.
\]
Hence $V$ is a solution to the perpetual American put option pricing problem (\ref{perpetual})--(\ref{perpetual-bc}), as claimed. 
\hfill$\diamondsuit$

\begin{remark}\label{rem-2}
In the case of a constant volatility function $\sigma(H)\equiv\sigma_0$ we have $\beta(u)=\frac{2}{\sigma_0^2} u$. It follows from equation (\ref{theo-rho}) that 

\[
 \varrho = E \frac{\gamma}{1+\gamma},\quad \hbox{where}\ \gamma=\frac{2 r}{\sigma_0^2},
\]
and,
\[
 V(S) =  \frac{S}{r} \int_0^{G^{-1}(-\ln(S/\varrho))} \frac{u}{u+r\beta(u)} d u
 =\frac{S}{r} \frac{1}{1+\gamma} G^{-1}(-\ln(S/\varrho)) = \frac{E}{1+\gamma} \left(\frac{S}{\varrho}\right)^{-\gamma}
\]
because $G(U)=\frac{1}{1+\gamma} \ln(U/U(x_0)), U(x_0)=r E/\varrho$, and so $G^{-1}(f) = \frac{r E}{\varrho} e^{(1+\gamma) f}$. Hence the solution is identical with Merton's explicit solution. 
\end{remark}

\subsection{Sensitivity analysis}

In this section we will investigate dependence of the free boundary position on model parameters. We consider the volatility function of the form:
\[
\frac12 \sigma(H)^2 H = \frac{\sigma_0^2}{2}\left(1 + \mu H^a\right) H + O(\mu^2) \quad \hbox{as}\ \mu\to0.
\]
Here $a\ge 0$ and $\mu\ge 0$ are specific model parameters. Our goal is to find the first order expansion of the free boundary position $\varrho$ considered as a function of a parameter $\mu$, i.e. $\varrho=\varrho(\mu)$. 

First, we derive expression for the derivative $\partial_\mu\beta$ of the inverse function $\beta$. For $H=\beta(u;\mu)$ we have $u= \frac12 \sigma(\beta(u;\mu))^2 \beta(u;\mu)$ and so
\[
0= \partial_\mu\left( \frac{\sigma_0^2}{2}\left(1 + \mu H^a\right) H\right)
= \frac{\sigma_0^2}{2}\left(1 + \mu (a+1) \beta^a\right)\partial_\mu\beta + \frac{\sigma_0^2}{2} \beta^{a+1} + O(\mu)
\]
For $\mu=0$ we have $\beta(u;0) = \frac{2}{\sigma_0^2}u$. Therefore
\[
 \partial_\mu\beta(u;0) = - (\sigma_0^2/2)^{-(a+1)} u^{a+1}.
\]
The first derivative of the free boundary position $\varrho=\varrho(\mu)$ can be deduced from the implicit equation (\ref{theo-rho}). We have 
\begin{eqnarray*}
0 &=& \frac{d}{d\mu} \int_0^{\frac{r E}{\varrho(\mu)}} \frac{\beta(u;\mu)}{u+r\beta(u;\mu)} d u 
\\
&=&  \frac{\beta(u;\mu)}{u+r\beta(u;\mu)}\Biggl|_{u=\frac{r E}{\varrho(\mu)}} \left(-\frac{r E}{\varrho(\mu)^2}\partial_\mu\varrho(\mu)\right)
+ \int_0^{\frac{r E}{\varrho(\mu)}} \frac{u \partial_\mu\beta(u;\mu)}{(u+r\beta(u;\mu))^2} d u . 
\end{eqnarray*}
Since, for $\mu=0$ we have $\varrho(0)= E\gamma/(1+\gamma)$ we conclude
\[
 \partial_\mu\varrho(0) = -\frac{E}{a+1}\gamma (1+\gamma)^{a-2}.
\]
In summary we have shown the following result:

\begin{theorem}\label{theo-sensitivity}
If the volatility function $\sigma(H)$ has the form $\frac12 \sigma(H)^2 H = \frac{\sigma_0^2}{2}\left(1 + \mu H^a\right) H + O(\mu^2)$ as $\mu\to0$, where $\mu,a\ge 0 $, then the free boundary position $\varrho=\varrho(\mu)$ of the perpetual American put option pricing problem has the asymptotic expansion:
\[
\varrho(\mu) = E\frac{\gamma}{1+\gamma} -  \mu \frac{E}{a+1}\frac{\gamma}{(1+\gamma)^{2-a} }+ O(\mu^2)\quad\hbox{as}\ \mu\to0. 
\]
\end{theorem}

\begin{remark}
In the case $a=0$ we have $\sigma(H)^2=\sigma_0^2(1+\mu)$. It corresponds to the constant volatility model. Thus $\varrho(\mu)=E\frac{\gamma(\mu)}{1+\gamma(\mu)}=E\frac{1}{1+1/\gamma(\mu)}$ where $\gamma(\mu)=2r/(\sigma_0^2(1+\mu))$. Hence
 \[
  \varrho(\mu)= E\frac{1}{1+\frac{\sigma_0^2}{2r}(1+\mu)}, \quad\hbox{and,}\ \ 
  \partial_\mu\varrho(0) =  - E\frac{\gamma}{(1+\gamma)^{2}},
 \]
as claimed by Theorem~\ref{theo-sensitivity}.
\end{remark}

\subsection{Comparison principle and Merton's sub-- and super--solutions}

In this section our aim is to derive sub-- and super--solutions to the perpetual American put option pricing problem. 

Let $\gamma>0$ be positive constant. By $V_\gamma$ we will denote the explicit Merton solution presented in Section~\ref{sec-merton}, i.e.

\begin{equation}
V_\gamma(S) = \left\{
\begin{array}{ll}
 \frac{E}{1+\gamma} \left(\frac{S}{\varrho_\gamma}\right)^{-\gamma}, & S> \varrho_\gamma, \\ 
 E-S,  &  0<S\le \varrho_\gamma,
\end{array} 
\right.
\label{explicit-solution-gamma}
\end{equation}
where
\begin{equation}
\varrho_\gamma = E \frac{\gamma}{1+\gamma}.
 \label{explicit-rho-gamma}
\end{equation}
It means that the pair $(V_\gamma(\cdot), \varrho_\gamma)$ is the explicit Merton solution corresponding to the constant volatility $\sigma_0^2=2r/\gamma$ (see (\ref{explicit-solution})).

Then, for the transformed function $U_\gamma(x)$ we have
\[
 U_\gamma(x) = -r S\partial_S \left(\frac{V_\gamma(S)}{S}\right) = r E \varrho_\gamma^\gamma e^{-(1+\gamma)x},\quad \hbox{for}\ 
 x=\ln S > x_{0\gamma} = \ln \varrho_\gamma.
\]
Clearly,
\begin{equation}
\partial_x U_\gamma + U_\gamma + r \beta(U_\gamma) = r \beta(U_\gamma) - \gamma U_\gamma.
\label{U-eq}
\end{equation}

Next we will construct a super--solution to the solution $U$ of the equation 
$\partial_x U_\gamma = - U_\gamma - r \beta(U_\gamma)$ by means of the Merton solution $U_\gamma$ where $\gamma=\gamma^+$ is the unique root of the equation
\begin{equation}
\gamma^+ \sigma(1+\gamma^+)^2 = 2r.
\label{gammaplus}
\end{equation}
Since 
\[
 U_{\gamma^+}(x) \le  U_{\gamma^+}(x_{0\gamma})= \frac{r E}{\gamma^+} = r\frac{1+\gamma^+}{\gamma^+}
\]
we obtain
\[
\frac12 \sigma((\gamma^+/r) U_{\gamma^+}(x))^2 \frac{\gamma^+}{r} U_{\gamma^+}(x)
\le \frac12 \sigma(1+\gamma^+)^2 \frac{\gamma^+}{r} U_{\gamma^+}(x).
\]
By taking the inverse function $\beta$ we finally obtain
\[
\frac{\gamma^+}{r} U_{\gamma^+}(x) \le \beta( U_{\gamma^+}(x) ).
\]
With regard to (\ref{U-eq}) we conclude that 
\begin{equation}
\partial_x U_{\gamma^+}(x) \ge - U_{\gamma^+}(x) - r \beta(U_{\gamma^+}(x))\quad\hbox{for}\ x> x_{0\gamma^+}=\ln\varrho_{\gamma^+}
\label{U-supersol}
\end{equation}

Similarly, we will construct the Merton sub--solution $U_{\gamma^-}$ satisfying the opposite differential inequality. Let $\gamma^-$ be given by
\begin{equation}
\gamma^- \sigma(0)^2 = 2r,
\label{gammaminus}
\end{equation}
i.e. $\gamma^-=2r/ \sigma(0)^2$. Then 
\[
 U_{\gamma^-} =\frac12 \sigma(0)^2 \frac{\gamma^-}{r} U_{\gamma^-}
 \le \frac12 \sigma( (\gamma^-/r) U_{\gamma^-})^2 \frac{\gamma^-}{r} U_{\gamma^-}
\]
and so, by taking the inverse function $\beta$ we obtain $\beta(U_{\gamma^-}) \le \frac{\gamma^-}{r} U_{\gamma^-}$. Then, from  (\ref{U-eq}) we conclude that
\begin{equation}
\partial_x U_{\gamma^-}(x) \le - U_{\gamma^-}(x) - r \beta(U_{\gamma^-}(x))\quad\hbox{for}\ x> x_{0\gamma^-}=\ln\varrho_{\gamma^-}.
\label{U-subsol}
\end{equation}

\begin{figure}
\begin{center}
\includegraphics[width=0.4\textwidth]{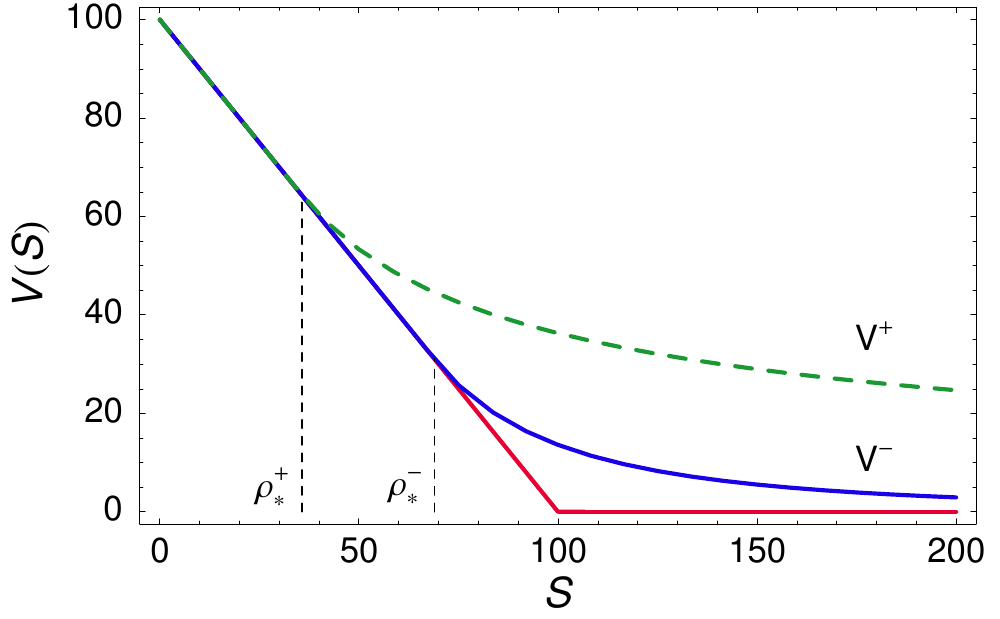}
\end{center}

\caption{A plot of Merton solutions $V^+$, $V^-$, and the pay-off diagram $\max(E-S,0)$ corresponding to constant volatilities $\sigma^+=0.6$ and $\sigma^-=0.3$ for model parameters: $E=100,  r=0.1$.}
\label{obr-amer-perpetual-bounds}
\end{figure}

In Fig.~\ref{obr-amer-perpetual-bounds} we plot Merton's solutions $V^\pm(\cdot)$ corresponding to  $\gamma^+ = 0.555$ ($\sigma^+\doteq 0.6$) and $\gamma^- = 2.222$ ($\sigma^-\doteq 0.3$) where $(\sigma^\pm)^2 = 2r/\gamma^\pm$. 

In what follows, we will prove the inequalities
\begin{equation}
\varrho_{\gamma^+} \le \varrho \le \varrho_{\gamma^-},
 \label{rhobounds}
\end{equation}
where $\varrho$ is the free boundary position for the nonlinear perpetual American put option pricing problem (\ref{perpetual})--(\ref{perpetual-bc}). 

Denote 
\[
 \beta^-(u) =\frac{\gamma^-}{r} u
\]
the inverse function to the function $H\mapsto \frac12 \sigma(0)^2 H$. As $\frac12 \sigma(0)^2 H \le \frac12 \sigma(H)^2 H$ we have $\beta(u)\le \beta^-(u)$ for any $u\ge0$.  Since
\[
 \int_0^{\frac{r E}{\varrho}} \frac{\beta(u)}{u+r\beta(u)} du = 1
 = \int_0^{\frac{r E}{\varrho_{\gamma^-}}} \frac{\beta^-(u)}{u+r\beta^-(u)} du
 \ge \int_0^{\frac{r E}{\varrho_{\gamma^-}}} \frac{\beta(u)}{u+r\beta(u)} du
\]
we conclude the inequality $\varrho \le \varrho_{\gamma^-}$. 

On the other hand, let 
\[
 \beta^+(u) =\frac{\gamma^+}{r} u
\]
be the inverse function to the function $H\mapsto \frac12 \sigma(1+\gamma^+)^2 H$. Then, 
for $u\le r E/\varrho_{\gamma ^+}$ we have
\[
 H=\beta(u) \le \beta(r E/\varrho_{\gamma ^+}) = \beta (\frac12 \sigma(1+\gamma^+)^2 (1+\gamma^+)) = 1+\gamma^+. 
\]
Therefore, for  $u\le r E/\varrho_{\gamma ^+}$ we have
$\beta(u) \ge \beta^+(u)$ and arguing similarly as before we obtain the estimate $\varrho_{\gamma^+} \le \varrho$ and so the inequalities (\ref{rhobounds}) follows.

For initial conditions we have $U_{\gamma^\pm}(x_{0\gamma^\pm}) = \frac{r E}{\varrho_{\gamma^\pm}}, U(x_0) = \frac{r E}{\varrho}$ and so
\[
 U_{\gamma^-}(x_{0\gamma^-})\le U(x_{0})\le U_{\gamma^+}(x_{0\gamma^+}).
\]
Using the comparison principle for solutions of ordinary differential inequalities we have $U_{\gamma^-}(x)\le U(x)\le U_{\gamma^+}(x)$. Taking into account the explicit form of the function $V(S)$ from Theorem~\ref{th-1} (see (\ref{theo-V})) we conclude the following result:

\begin{theorem}\label{theo-comparison}
Let $(V(\cdot),\varrho)$ be the solution to the perpetual American pricing problem  (\ref{perpetual})--(\ref{perpetual-bc}). Then 
\[
 V_{\gamma^-}(S) \le V(S) \le V_{\gamma^+}(S) \quad\hbox{for any}\ \ S\ge 0,
\]
and, 
\[
 \varrho_{\gamma^+} \le \varrho \le \varrho_{\gamma^-}
\]
where $V_{\gamma^\pm}, \varrho_{\gamma^\pm}$ are explicit Merton's solutions where $\gamma^\pm$ are given by (\ref{gammaplus}) and (\ref{gammaminus}).
\end{theorem}

A graphical illustration of the comparison principle is shown in Fig.~\ref{obr-FreyRAPM-perpetual}.

\section{Numerical approximation scheme and computational results}

In this section we propose a simple and efficient numerical scheme for constructing a solution to the perpetual put option problem (\ref{perpetual})--(\ref{perpetual-bc}). 

Using transformation $H=\beta(u)$, i.e. $u=\frac12\sigma(H)^2 H$ and $du=\frac12 \partial_H(\sigma(H)^2 H) dH$ we can rewrite the equation for the free boundary position (see (\ref{theo-rho})) as follows:
\begin{equation}
\int_0^{\beta(r E/\varrho)} \frac{H \frac12 \partial_H(\sigma(H)^2 H) }{\frac12\sigma(H)^2 H +r H} d H =1.
\label{theo-rho-mod}
\end{equation}
Similarly, the option price (\ref{theo-rho}) can be rewritten in terms of the $H$ variable as follows:
\begin{equation}
V(S) = \frac{S}{r} \int_0^{\beta(G^{-1}(-\ln(S/\varrho)))} 
\frac{\frac12\sigma(H)^2 H \frac12 \partial_H(\sigma(H)^2 H) }{\frac12\sigma(H)^2 H +r H} d H.
\label{theo-V-mod}
\end{equation}
With this transformation we can reduce computational complexity in the case when the inverse function $\beta$ is not given by a closed form formula. 

\subsection{Numerical results}

Results of numerical calculation for the Frey model (\ref{doplnky-c-frey}) and the RAPM model (\ref{doplnky-c-RAPM}) are summarized in Tables~\ref{tab2} and \ref{tab3}. We show the position of the free boundary $\varrho$ and the perpetual option value $V$ evaluated at the exercise price $S=E$. The results are computed for various values of the parameter $\mu$ for the Frey model and the RAPM model. Other model parameter were chosen as: $E=100,  r=0.1$ and $\sigma_0=0.3$.

\begin{figure}
\begin{center}
\includegraphics[width=0.4\textwidth]{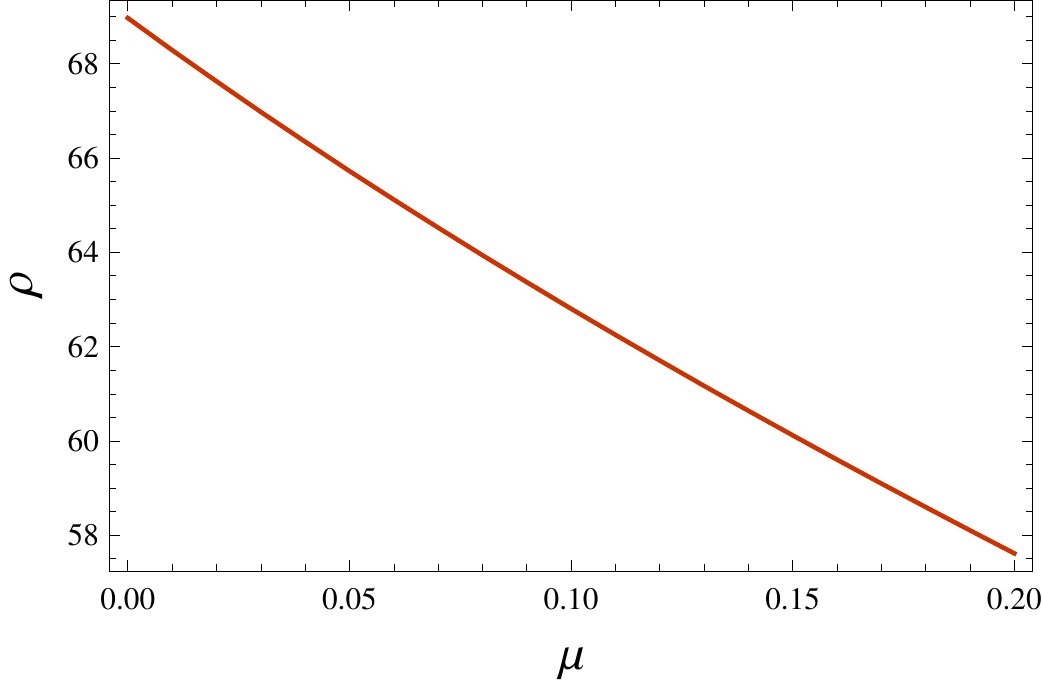}
\qquad 
\includegraphics[width=0.4\textwidth]{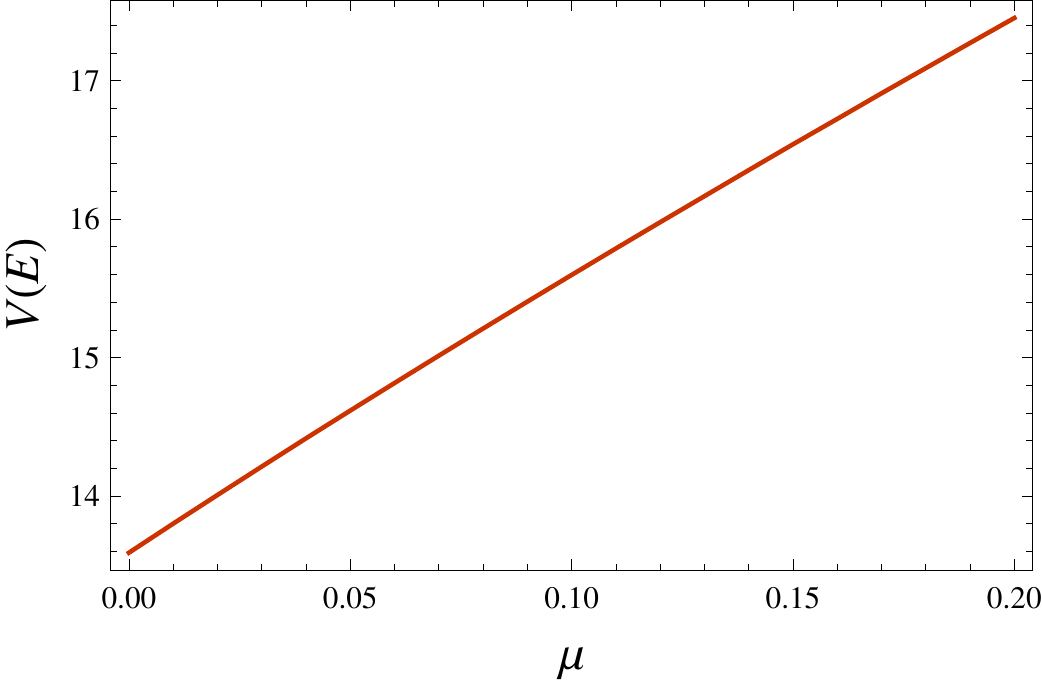}

\quad a) \hskip 6truecm b)

\end{center}

\caption{A plot of dependence of the free boundary position $\varrho$ a) and the perpetual American put option price $V(E)$ b) on the model parameter $\mu$ for the Frey model (\ref{doplnky-c-frey}).}
\label{obr-sensitivity-Frey}
\end{figure}

\begin{table}
\caption{\small The free boundary position $\varrho=\varrho(\mu)$ and the option price $V(S)$ evaluated at $S=E$ for various values of the model parameter $\mu\ge0$ for the Frey model (\ref{doplnky-c-frey}).}
\label{tab2}

\begin{center}

\begin{tabular}
{c|ccccccc}
$\mu$& 
0.00& 
0.01& 
0.05& 
0.10& 
0.15& 
0.20& 
0.22 
\\
\hline\hline
$\varrho$& 
68.9655& 
68.2852& 
65.7246& 
62.8036& 
60.1175& 
57.6177& 
56.6627
\\
$V(E)$& 
13.5909& 
13.8005& 
14.6167& 
15.5961& 
16.5389& 
17.4510& 
17.8083 \\
\hline
\end{tabular}

\end{center}

\end{table}

In the Frey model  (\ref{doplnky-c-frey}) the nonlinear volatility function has the form:
\[
\sigma(H)^2 = \sigma_0^2 \left(1-\mu H\right)^{-2}.
\]
The range of the parameter $\mu$ is therefore limited to satisfy the strict inequality $1-\mu H = 1-\mu S\partial_S^2 V(S) >0$. However, using the identity
\[
 \frac{1}{1-\mu H} = 1 + \sum_{n=1}^\infty \mu^n H^n.
\]
we can approximate the Frey volatility function as follows:
\begin{equation}
\sigma(H)^2 = \sigma_0^2 \left(1 + \sum_{n=1}^N \mu^n H^n   \right)^{2},
\label{frey-modif}
\end{equation}
where $N$ is sufficiently large. Interestingly, a similar power series expansion of $\sigma(H)^2$ can be found in the generalized Black-Scholes model proposed by Cetin, Jarrow and Protter in \cite{Cetin2004}.

\begin{figure}
\begin{center}
\includegraphics[width=0.4\textwidth]{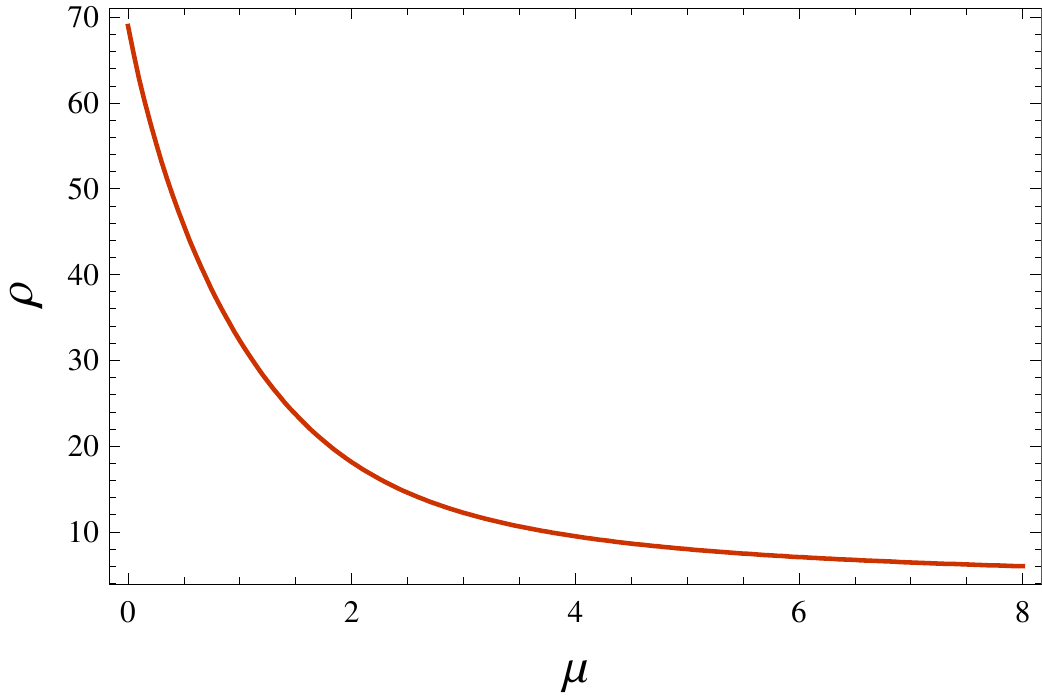}
\qquad 
\includegraphics[width=0.4\textwidth]{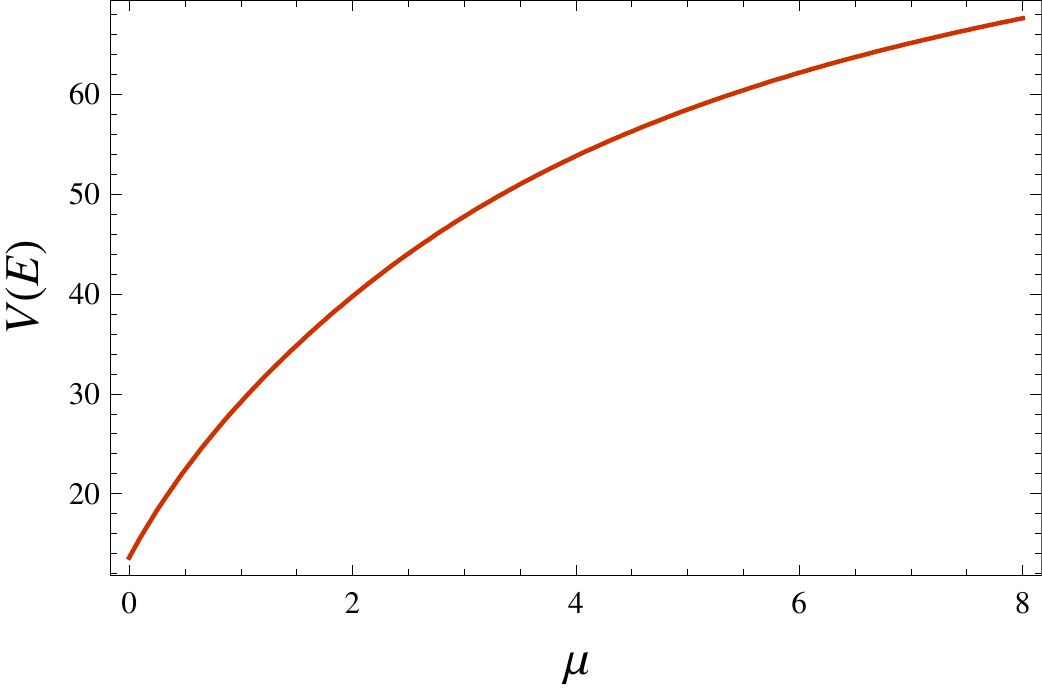}

\quad a) \hskip 6truecm b)

\end{center}

\caption{A plot of dependence of the free boundary position $\varrho$ a) and the perpetual American put option price $V(E)$ b) on the model parameter $\mu$ for the modified Frey model (\ref{frey-modif}).}
\label{obr-sensitivity-Frey2}
\end{figure}

In computations shown in Fig.~\ref{obr-sensitivity-Frey2} and Tab.~\ref{tab2b} we present results of the free boundary position and the perpetual American put option price $V(E)$ for $N=10$ and larger interval of parameter values $\mu\in[0,8]$. Note that the results for small values $\mu\le 0.1$ computed from the original Frey volatility (\ref{doplnky-c-frey}) and (\ref{frey-modif}) are very close to each other.

\begin{table}
\caption{The free boundary position $\varrho$ and the option price $V(S)$ evaluated at $S=E$ for various values of the model parameter $\mu\ge0$ for the modified Frey model.}
\label{tab2b}

\begin{center}
\begin{tabular}
{c|ccccccc}
$\mu$& 
0.00& 
0.10& 
0.50& 
1.00& 
2.00& 
4.00& 
8.00 \\
\hline\hline
$\varrho$& 
68.9655& 
62.8037& 
45.3007& 
31.0862& 
16.3126& 
8.3818& 
5.4556
\\
$V(E)$& 
13.5909& 
15.5961& 
22.4529& 
29.5719& 
41.0654& 
56.1777& 
70.2259 \\
\hline
\end{tabular}
\end{center}

\end{table}

\begin{figure}
\begin{center}
\includegraphics[width=0.4\textwidth]{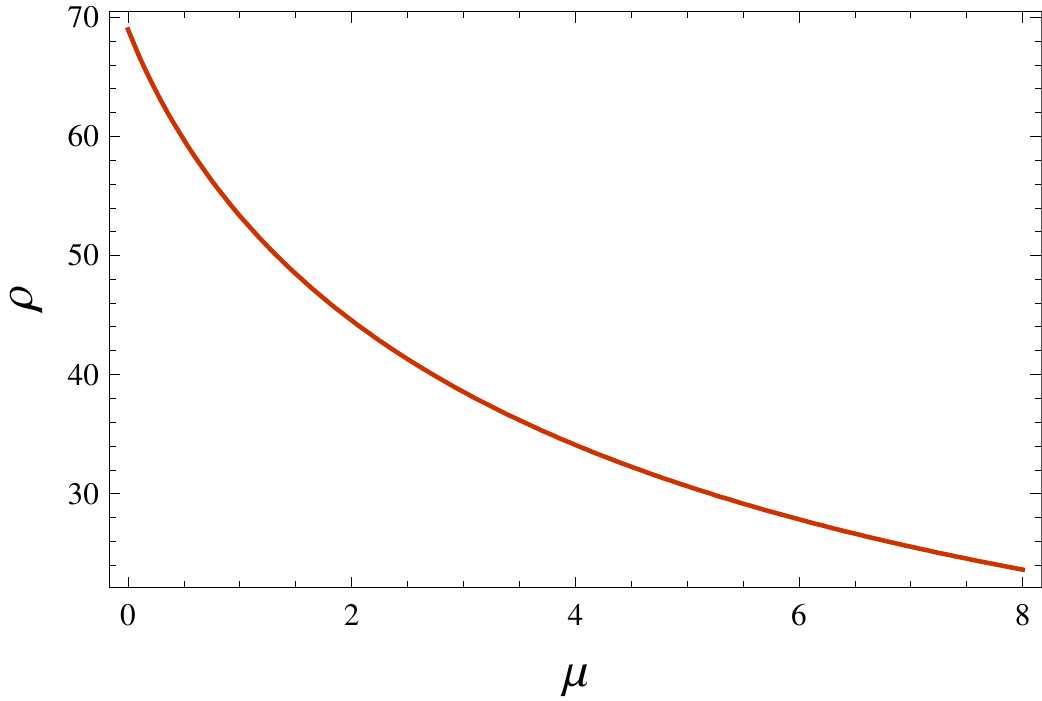}
\qquad 
\includegraphics[width=0.4\textwidth]{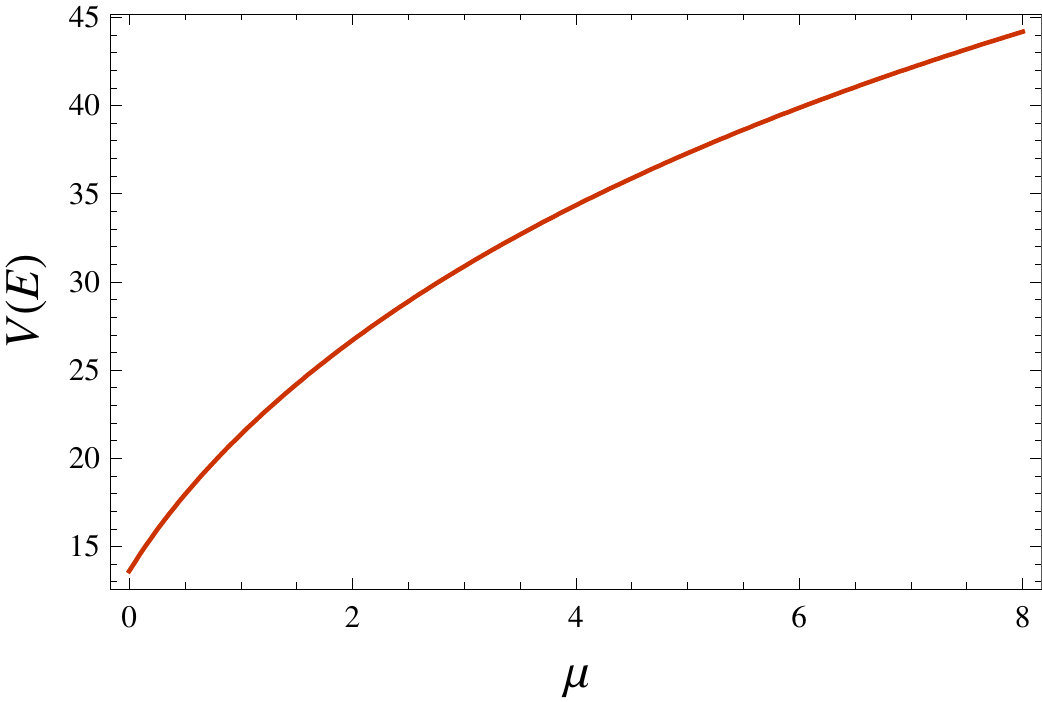}

\quad a) \hskip 6truecm b)

\smallskip

\includegraphics[width=0.4\textwidth]{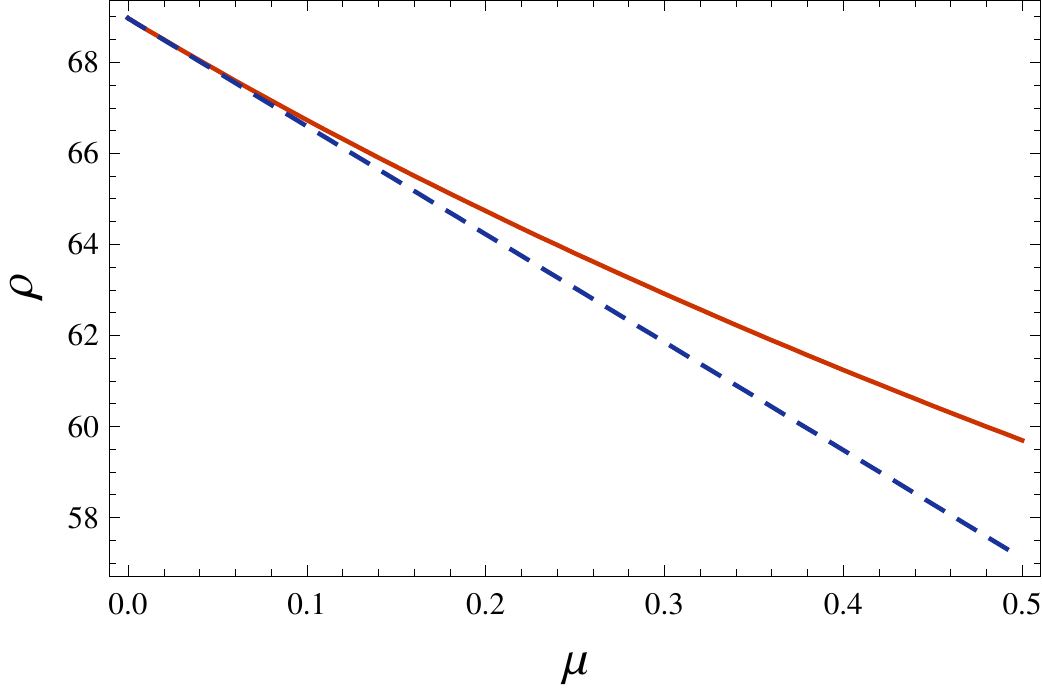}

c)
\end{center}

\caption{A plot of dependence of the free boundary position $\varrho$ a) and the perpetual American put option price $V(E)$ b) on the model parameter $\mu$ for the RAPM model (\ref{doplnky-c-RAPM}). The comparison of the free boundary position and its linear approximation c).}
\label{obr-sensitivity-RAPM}
\end{figure}

In our next computational example we consider the Risk adjusted pricing methodology model (RAPM). In computations shown in Fig.~\ref{obr-sensitivity-RAPM}, a) and Tab.~\ref{tab3} we present results of the free boundary position and the perpetual American put option price $V(E)$ for the RAPM model (see Fig.~\ref{obr-sensitivity-RAPM}, b)). We also show comparison of the free boundary position $\varrho=\varrho(\mu)$ and its linear approximation derived in Theorem~\ref{theo-sensitivity} (see Fig.~\ref{obr-sensitivity-RAPM}, c)). 

\begin{table}
\caption{The free boundary position $\varrho$ and the option price $V(S)$ evaluated at $S=E$ for various values of the model parameter $\mu\ge0$ for the RAPM model.}
\label{tab3}

\begin{center}

\begin{tabular}
{c|ccccccc}
$\mu$& 
0.00& 
0.10& 
0.50& 
1.00& 
2.00& 
4.00& 
8.00 \\
\hline\hline
$\varrho$& 
68.9655& 
66.7331& 
59.6973& 
53.3234& 
44.5408& 
34.0899& 
23.6125 \\
$V(E)$& 
13.5909& 
14.5761& 
17.9398& 
21.3434& 
26.6857& 
34.3393& 
44.1774 \\
\hline
\end{tabular}

\end{center}

\end{table}

\begin{figure}
\begin{center}
\includegraphics[width=0.4\textwidth]{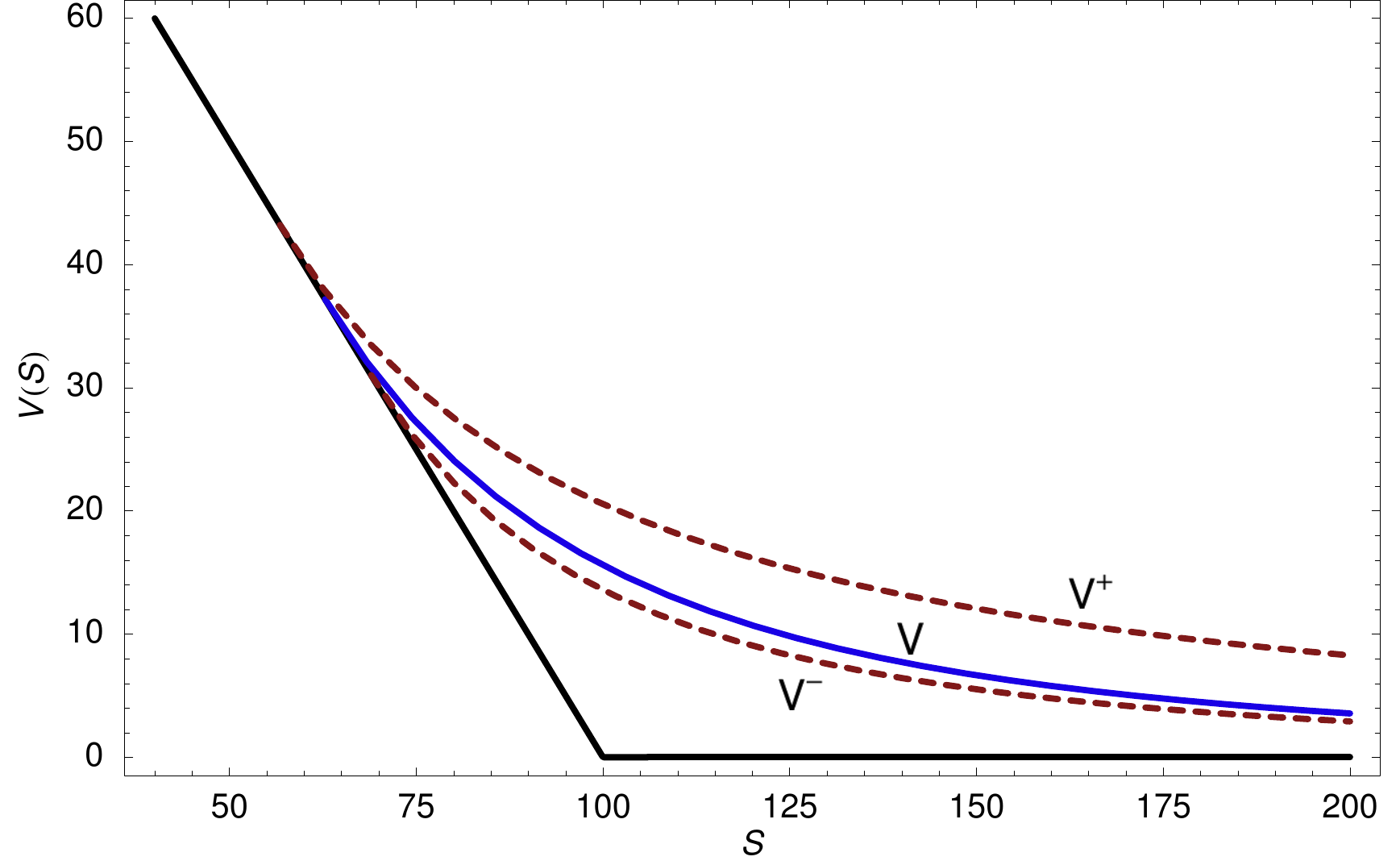}
\qquad 
\includegraphics[width=0.4\textwidth]{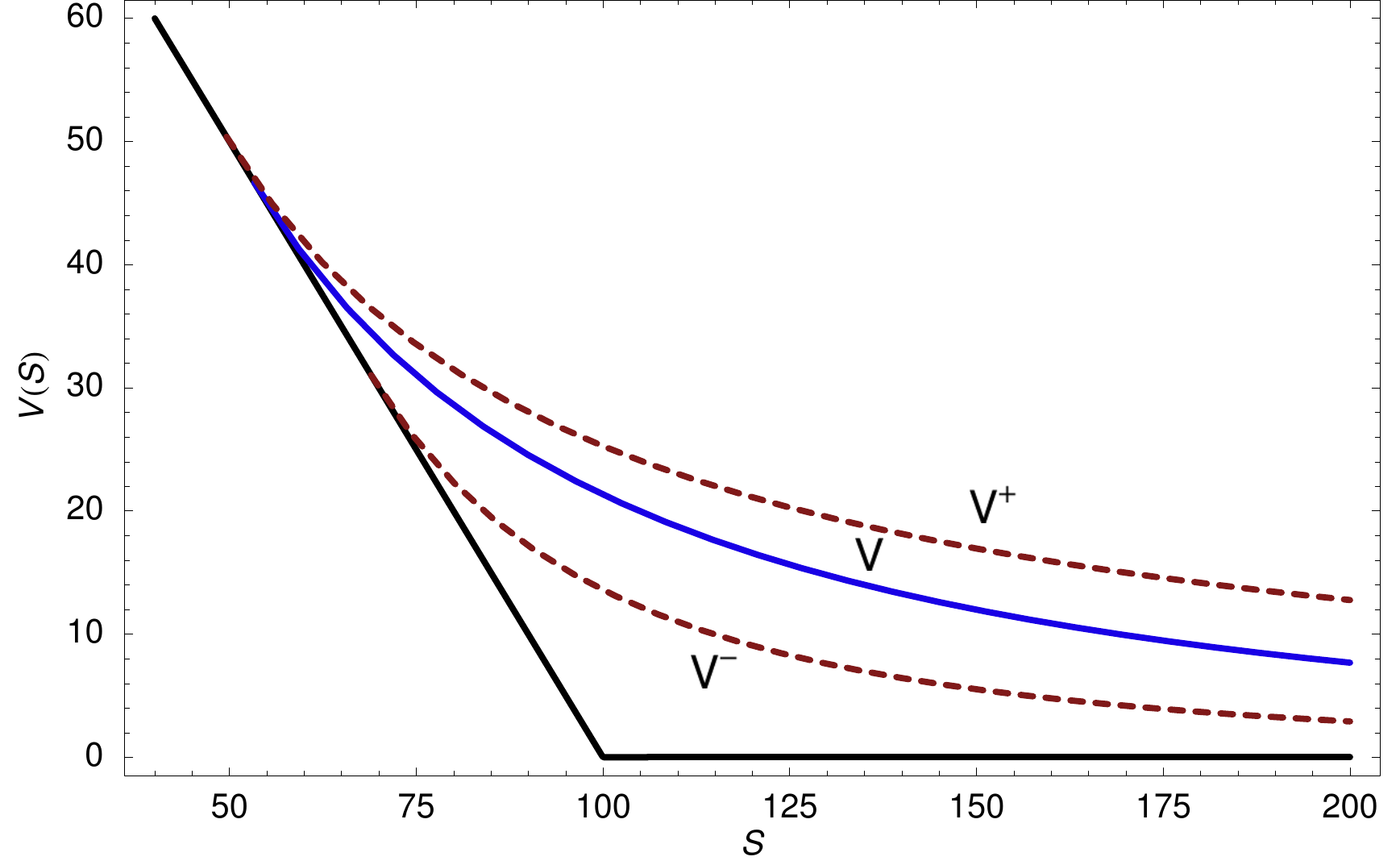}

\quad a) \hskip 6truecm b)
\end{center}

\caption{The solid curve represents a graph of a perpetual American put option $S\mapsto V(S)$ for the Frey model a) with $\mu=0.1$ and the RAPM model b) with $\mu=1$. Sub- and super- solutions $V^-=V_{\gamma^-}$ and $V^+=V_{\gamma^+}$ are depicted by dashed curves, $V^+<V^-$. The model parameters: $E=100,  r=0.1$ and $\sigma_0=0.3$.}
\label{obr-FreyRAPM-perpetual}
\end{figure}

In the last examples shown in Fig.~\ref{obr-FreyRAPM-perpetual} we present comparison of the option price $V(S)$ and the free boundary position $\varrho$ for the Frey model (left) and the Risk adjusted pricing methodology model (right) with closed form explicit Merton's solutions corresponding to the constant volatility.

\section{Conclusions}
In this paper we analyzed the free boundary problem for pricing perpetual American put option when the volatility is a function of the second derivative of the option price. We showed how the problem can be transformed into a single implicit equation for the free boundary position and explicit integral expression for the option price.

\section*{Acknowledgements}
This research was supported by the European Union in the FP7-PEOPLE-2012-ITN project STRIKE - Novel Methods in Computational Finance (304617), the project CEMAPRE – MULTI/00491 financed by FCT/MEC through national funds and the Slovak research Agency Project VEGA 1/0251/16.

\section*{Competing interests}
The authors declare that they have no competing interests.

\end{document}